\documentclass[aps,prl,nofootinbib,showpacs,floatfix,twocolumn]{revtex4}  
\usepackage{bm}
\usepackage{latexsym}
\usepackage{dcolumn}
\usepackage{amsfonts,amssymb,amsmath}
\usepackage{graphicx,epsfig}
\usepackage{psfrag}
\usepackage{subfigure}
\usepackage{url}
\usepackage{feynmf}
\usepackage{rotating}
\usepackage{hyperref}
\usepackage{enumitem}
\usepackage{color}
\usepackage[export]{adjustbox}
\hypersetup{
    bookmarks=true,         
    unicode=false,          
    pdftoolbar=true,        
    pdfmenubar=true,        
    pdffitwindow=false,     
    pdfstartview={FitH},    
    pdftitle={My title},    
    pdfauthor={Author},     
    pdfsubject={Subject},   
    pdfcreator={Creator},   
    pdfproducer={Producer}, 
    pdfkeywords={keyword1} {key2} {key3}, 
    pdfnewwindow=true,      
    colorlinks=true,       
    linkcolor=red,          
    citecolor=cyan,        
    filecolor=magenta,      
    urlcolor=green,           
    linktocpage=true
}

\def\beq{\begin{equation}}
\def\eeq{\end{equation}}
\def\br{\begin{eqnarray}}
\def\er{\end{eqnarray}}
\def\benu{\begin{enumerate}}
\def\efnu{\end{enumerate}}

\def\l{\left}
\def\r{\right}

\begin{document}




\title{Joining bits and pieces of reionization history}
\author{Dhiraj Kumar Hazra} \email{dhiraj@imsc.res.in}
\affiliation{The Institute of Mathematical Sciences, HBNI, CIT Campus, Chennai 600113, India\\INAF OAS Bologna, Osservatorio di Astrofisica e Scienza dello Spazio di Bologna, via Gobetti 101, I-40129 Bologna, Italy\\INFN, Sezione di Bologna, via Irnerio 46, 40126 Bologna, Italy}
\author{Daniela Paoletti} \email{daniela.paoletti@inaf.it}
\affiliation{INAF OAS Bologna, Osservatorio di Astrofisica e Scienza dello Spazio di Bologna, via Gobetti 101, I-40129 Bologna, Italy\\INFN, Sezione di Bologna, via Irnerio 46, 40126 Bologna, Italy}
\author{Fabio Finelli} \email{fabio.finelli@inaf.it}
\affiliation{INAF OAS Bologna, Osservatorio di Astrofisica e Scienza dello Spazio di Bologna, via Gobetti 101, I-40129 Bologna, Italy\\INFN, Sezione di Bologna, via Irnerio 46, 40126 Bologna, Italy}
\author{George F. Smoot}\email{gfsmoot@lbl.gov}
\affiliation{Paris Centre for Cosmological Physics, Universit\'e ́de Paris,
CNRS, Astroparticule et Cosmologie, F-75013 Paris, France;\\
Institute for Advanced Study \& Physics Department, Hong Kong University of Science and Technology, Clear Water Bay, Kowloon, Hong Kong;\\
Physics Department and Lawrence Berkeley National Laboratory, University of California, Berkeley, CA 94720, USA;\\
Energetic Cosmos Laboratory, Nazarbayev University, Astana, Kazakhstan}
\date{\today}

\begin{abstract}
Cosmic Microwave Background (CMB) temperature and polarization anisotropies from Planck have estimated a lower value of the optical depth 
to reionization ($\tau$) compared to WMAP. A significant period in the reionization history would then fall within $6<{\rm redshift} (z)< 10$,  
where detection of galaxies with Hubble Frontier Fields (HFF) 
program and independent estimation of neutral hydrogen in the inter galactic medium by Lyman-$\alpha$ observations are also available. This overlap allows an analysis of cosmic reionization which utilizes a direct combination of CMB and these astrophysical measurements and potentially breaks degeneracies in parameters describing the physics of reionization.
For the first time we reconstruct reionization histories by assuming photo-ionization and recombination rates to 
be free-form and by allowing underlying cosmological parameters to vary with CMB (temperature and polarization anisotropies and lensing) data from Planck 2018 release and a compilation of astrophysical data. 
We find an excellent agreement between the low-$\ell$ Planck 2018 HFI polarization likelihood and astrophysical data in determining the integrated optical depth. By combining both data, 
we report for a minimal reconstruction $\tau=0.051^{+0.001+0.002}_{-0.0012-0.002}$ at 68\% and 95\% CL, which, for the errors in the current astrophysical
measurements quoted in the literature, is nearly twice better than the
projected cosmic variance limited CMB measurements. For the duration of reionization, redshift interval between 10\% and complete ionization, we get 
$2.9^{+0.12+0.29}_{-0.16-0.26}$ at 68\% and 95\% CL, which improves significantly on the corresponding result obtained by using Planck 2015 data. By a Bayesian analysis of the combined results we do not find evidence beyond monotonic reionization histories, therefore multi-phase reionization scenario such as a first burst of reionization followed by recombination plateau and thereafter complete reionization is disfavored compared to minimal alternatives.

\end{abstract}

\maketitle

\section{Introduction}

The two cosmic transitions between ionized and neutral state for the
hydrogen atom are imprinted in key astrophysical and cosmological observations.
The first transition from ionized plasma to neutral state for atoms, {\it cosmological recombination}
occurred around 13.8 billion of years ago (or equivalently at a redshift $z\sim 1100$).
After half a billion years, the hydrogen became ionized again during {\it cosmic reionization}, which followed the so called dark ages.
Evidence for cosmic reionization comes from astrophysical measurements, such as the Gunn-Peterson test in high redshift quasars
or the declining visibility of Lyman-$\alpha$ high redshift galaxies,
and from cosmological observations as the large angular scale polarization pattern of CMB anisotropies.
While astrophysical measurements mostly encode the central stage and the
completion of cosmic reionization, the CMB anisotropy pattern is mostly sensitive to its duration
through the integrated optical depth ($\tau$), and marginally to its early stage.

Recent determinations of $\tau$ from Planck assuming a nearly instantaneous transition
for the ionization fraction~\cite{Planck2018:param,Planck2018:like,Planck2016:reion,Planck2016:HFIsys,Lattanzi2016}
have revealed preferences for lower values compared to WMAP, owing to the 
understanding of the Galactic dust contamination to microwave polarization at large angular scales.
Recent works demonstrate that star forming galaxies detected till $z\simeq10$ as a source of reionization offer 
a consistent scenario with this optical depth~\cite{Ishigaki2018}.
Observation of galaxies at high redshifts ($z\sim6-10$), mainly with recent
six cluster observations (Abell 2744, MACSJ0416, MACSJ0717, MACSJ1149, AbellS1063, Abell370)
by Hubble Frontier Fields program~\cite{Coe2015,Lotz2017,Livermore2017} upto a limiting AB magnitude of 
29, provides shape of UV luminosity densities that determine the ionizing photon emission history. 
On the other hand, Gunn-Peterson optical depth~\cite{Fan2006:annurev,Fan2006}; ionized near-zone around high redshift quasars~\cite{Mortlock2011,Bolton2011}; dark gaps in quasar spectra~\cite{McGreer:2014qwa}; damping wings of Gamma-Ray Burst 050904~\cite{Totani:2005ng,McQuinn:2007gm} and quasars~\cite{Schroeder2013,Greig:2016vpu,Davies:2018pdw}; Lyman-$\alpha$ emitters~\cite{McQuinn:2007dy,Ouchi2010}, Lyman-$\alpha$ emission from galaxies~\cite{Ono2012,Caruana2013,Schenker2014,Tilvi2014,Pentericci2014,Sobacchi2015,Mason:2017eqr,Mason:2019ixe} provide measurement of remaining neutral hydrogen in the inter-galactic medium (IGM) between redshift $5-8$. 
Redshift overlap of HFF and Lyman-$\alpha$ 
observations with the reionization as measured by Planck
calls for a joint analysis in a model independent framework. 
Since physics describing cosmic reionization is partially degenerate with cosmological parameters~\cite{Zaldarriaga:1997ch,HPFS18}, 
it is important to perform this analysis by allowing the underlying cosmological model to vary as 
well (see~\cite{Robertson2010:Nature,Mitra2011,Mitra2015,Bouwens2015,Robertson2015,Ishigaki2015,Price2016,Gorce2017,Ishigaki2018,Mitra2018} for previous works in which all but the reionization parameters are kept fixed).

In this Letter we perform for the first time a joint analysis using updated CMB anisotropy 
and a combination of astrophysical data to reconstruct reionization histories, where solutions to ionization equation of hydrogen with free-form ionization and recombination rates are used instead of conventional free-electron fraction parametrization~\cite{Hu2003,Mortonson2007,Douspis2015,HS17,HPFS18,HPBFSSS,Obied:2018qdr,Heinrich2018,Villanueva-Domingo:2017ahx,Millea:2018bko}. 
Our analysis removes parametric model dependence with this generic construct. Use of the ionization equation allows us to include all three types of data (CMB, UV luminosities and neutral fraction) in a single framework. At the same time use of complete CMB data and freedom in the cosmological parameters exploits the degeneracies and provide conservative constraints.

\section{Reconstruction of reionization history: the framework}
We directly solve the reionization equation~\cite{Wyithe:2002qu} for the volume filling factor of ionized regions:
\beq
\frac{dQ_{\rm HII}}{dt}=\frac{\dot{n}_{\rm ion}}{\langle n_{\rm H}\rangle}-\frac{Q_{\rm HII}}{t_{\rm rec}},\,
\eeq
where the source term $\dot{n}_{\rm ion}$ is the ionizing photon production rate and is defined by the product of the UV luminosity density ($\rho_{\rm UV}$),
the photon production efficiency ($\xi_{\rm ion}$) and the escape fraction ($f_{\rm esc}$). We keep the magnitude averaged product 
$\log_{10}\langle f_{\rm esc}\xi_{\rm ion}\rangle=24.85$ from~\cite{Ishigaki2015}, also consistent with other analyses~\cite{Price2016,Ishigaki2018,Madau2017}. 
The recombination time is defined as $t_{\rm rec}=1/\l[C_{\rm HII}\alpha_{\rm B}(T) (1+Y_p/(4X_p))\langle n_{\rm H}\rangle (1+z)^3\r]$ using the clumping factor ($C_{\rm HII}$), recombination coefficient ($\alpha_{\rm B}(T)$), density of hydrogen 
atom ($\langle n_{\rm H}\rangle$) and hydrogen ($X_p$) and helium abundances ($Y_p$). In this work, instead of 
using analytical forms, we define $\rho_{\rm UV}$ and $t_{\rm rec}$ to be free parameters in
different nodes which are allowed to vary between a conservative redshift range for the reionization process, i.e. $z=5.5-30$ (see,~\cite{HS17}).
Different nodes are connected using Piecewise Cubic Hermite Interpolating Polynomial. Fixed nodes are located at $z=0,5.5$
and 30 and values of source and recombination terms are fixed to be consistent to best fit logarithmic double
power law (see Eq. (39) of~\cite{Ishigaki2015}) and also consistent with~\cite{Becker2013} when interpolated at smaller redshifts. However as we allow the intermediate source and recombination terms to be free, values at fixed nodes do not limit our general construct. 

We allow upto three nodes in this moving-bin reconstruction denoted as $B1,~B2$ and $B3$ respectively and each node comes with three parameters, namely, the intermediate redshift ($z_{\rm int}$) and $\rho_{\rm UV}$ and recombination timescale defined 
in that redshift. Since reionization progresses with the competition between ionization and recombination, in our analysis we have used the ratios $(1/t_{\rm rec})/(\dot{n}_{\rm ion}/\langle n_{\rm H}\rangle)$ as free parameter instead of $t_{\rm rec}$. From the reconstruction, $t_{\rm rec}$ can be obtained as a function of redshift and assuming certain IGM temperature, the clumping factor can also be derived. For a minimal construct we also consider $B0$ where we impose at $z=z_{\rm int}$, $\dot{n}_{\rm ion}t_{\rm rec}=\langle n_{\rm H}\rangle$ (a mathematical limit of B1 that generalizes the ionization balance).
The optical depth is a derived parameter in our approach and is given by the integral from onset of reionization ($z_{\rm begin}$) till today: $\tau=\int_{0}^{z_{\rm begin}}\frac{c(1+z)^2}{H(z)}Q_{\rm HII}(z)\sigma_{\rm Thomson} \langle n_{\rm H}\rangle(1+\frac{Y_p}{4X_p})$,
where $\sigma_{\rm Thomson}$ is the Thomson scattering cross-section. We fuse our integrator with {\tt CAMB}~\cite{Lewis1999} maintaining the standard treatment for Helium reionization.  

\section{Datasets and priors}~\label{sec:data} 

Three different datasets have been mostly used in this work. For CMB we use the latest publicly available likelihoods in temperature, polarization 
and lensing from the Planck 2018 release (hereafter P18)~\cite{Planck2018:like,Planck2018:Lensing} and the Planck 2015 release (hereafter P15) ~\cite{Planck2015:like,Planck2015:Lensing}. 
We use the full angular power spectrum data 
in order to fully account for non-negligible correlations between reionization history and other cosmological parameters~\cite{HPFS18}.
For UV luminosity density, we use~\cite{Bouwens2014,Ishigaki2018} data spanning $z\sim6-11$ derived from Hubble Frontier Fields~\cite{HFFsite,Lotz2017} observations. The density is obtained by integrating the UV luminosity function by fitting Schechter function 
till a truncation magnitude of $-17$ (hereafter UV17) [We use the recent data compiled by Ishigaki {\it et. al} (2018) exploiting the full six-cluster HFF data]. We also use direct $Q_{\rm HII}$ constraints (hereafter QHII) from~\cite{McGreer:2014qwa,Totani:2005ng,McQuinn:2007gm,Schroeder2013,Greig:2016vpu,Davies:2018pdw,McQuinn:2007dy,Ouchi2010,Schenker2014,Tilvi2014,Mason:2017eqr,Mason:2019ixe}. These data cover a redshift range of $5.6-8$ and thereby overlap with the UV density. For B0 and B1 the intermediate redshift 
is allowed to vary between the entire range $z=5.5-30$. For B2, $z_{\rm int}^1$ can vary between $z=5.5-12$  
and $z_{\rm int}^2$ between $z=12-30$; for B3, $z_{\rm int}^1$, $z_{\rm int}^2$ and $z_{\rm int}^3$ move within $5.5-8$, $8-12$ and 
$12-30$ respectively. The redshift ranges for nodes in B2 and B3 are chosen in a way that the UV data can constrain the source term variation till $z=12$ and the CMB constrains the last node by constraining the integrated optical depth. 
We allow $\Omega_{\rm b}h^2$, $\Omega_{\rm CDM}h^2$, $\theta$, $A_{\rm s}$, $n_{\rm s}$, foregrounds and calibration nuisance 
parameters in Planck likelihood to vary. We use publicly available {\tt CosmoMC}~\cite{Lewis2002} for parameter estimation.
We also consider the stability of our results allowing $\log_{10}\langle f_{\rm esc}\xi_{\rm ion}\rangle$ to vary and using UV luminosity data with truncation magnitude of $-15$.

\section{Constraints and concordances}~\label{sec:results}

We first show the consistency of the low-$\ell$ Planck 2018 polarisation likelihood lowE and astrophysical data UV17+QHII in determining the integrated optical depth $\tau$.
By combining Planck 2018 TT and astrophysical data we obtain for B0, $\tau = 0.051 \pm 0.001$ at 68 \%CL. This determination of $\tau$ is in excellent agreement, but much more precise, than the 68 \% CL estimate obtained by
Planck 2018 TT + lowE, $\tau = 0.052 \pm 0.008$
The consistency of Planck lowE and astrophysical data in determining $\tau$ is robust to the addition of Planck high-$\ell$ polarisation and/or lensing and to the addition of nodes in the rates.

We now proceed with the joint constraints using P15, P18, P18+UV17 and P18+UV17+QHII in minimal single node (B0), single node (B1), 2 nodes (B2) cases and P15+UV17, P18+UV17 and P18+UV17+QHII in three nodes (B3) case. For B3 we do not explore CMB-only constraints owing to its inability to provide reasonable constraints in such an extended parameter space.
\begin{table}[!htb]
\begin{tabular}{|l|l|l|l|l|}
\hline
\multicolumn{2}{|l|}{Model/data}    & P18 & P18+UV17 & P18+UV17+QHII \\ \hline
\begin{tabular}[c]{@{}l@{}}B0\\ Minimal\end{tabular} & \begin{tabular}[c]{@{}l@{}}$\chi^2_\mathrm{eff}$\\ $\ln B$\\ $\tau$\\ $\Delta^\mathrm{reion}_z$\end{tabular} 
& \begin{tabular}[c]{@{}l@{}} 2779.9\\0\\$0.051^{+0.006}_{-0.009}$\\$3^{+0.79+2.0}_{-1.2-1.8}$\end{tabular}   
& \begin{tabular}[c]{@{}l@{}} 2783.4\\0\\$0.05\pm0.001$\\$2.8^{+0.11+0.27}_{-0.15-0.25}$\end{tabular}          
& \begin{tabular}[c]{@{}l@{}} 2792\\0\\$0.051\pm0.001$\\$2.9^{+0.12+0.29}_{-0.16-0.26}$\end{tabular}               \\ \hline
\begin{tabular}[c]{@{}l@{}}B1\\ 1 node\end{tabular}  & \begin{tabular}[c]{@{}l@{}}$\chi^2_\mathrm{eff}$\\ $\ln B$\\ $\tau$\\ $\Delta^\mathrm{reion}_z$\end{tabular} 
& \begin{tabular}[c]{@{}l@{}} 2780.5\\-0.4\\$0.052^{+0.006}_{-0.009}$\\$3.08^{+0.77+2.1}_{-1.3-2.0}$\end{tabular}   
& \begin{tabular}[c]{@{}l@{}} 2782\\-0.1\\$0.05\pm0.001$\\$2.8^{+0.11+0.31}_{-0.15-0.24}$\end{tabular}          
& \begin{tabular}[c]{@{}l@{}} 2790.3\\0\\$0.051\pm0.001$\\$2.9^{+0.12+0.29}_{-0.16-0.26}$\end{tabular}               
\\ \hline
\begin{tabular}[c]{@{}l@{}}B2\\ 2 nodes\end{tabular} & \begin{tabular}[c]{@{}l@{}}$\chi^2_\mathrm{eff}$\\ $\ln B$\\ $\tau$\\ $\Delta^\mathrm{reion}_z$\end{tabular}
& \begin{tabular}[c]{@{}l@{}} 2778.8\\-2.2\\$0.05\pm0.008$\\$3.3^{+0.03+7}_{-2.7-3}$\end{tabular}   
& \begin{tabular}[c]{@{}l@{}} 2782\\-3.5\\$0.049^{+0.007}_{-0.006}$\\$2.7^{+0.2+1.3}_{-0.32-0.8}$\end{tabular}          
& \begin{tabular}[c]{@{}l@{}} 2789\\-3.2\\$0.052^{+0.0008}_{-0.002}$\\$3.05^{+0.08+1.2}_{-0.53-0.7}$\end{tabular}               
\\ \hline
\begin{tabular}[c]{@{}l@{}}B3\\ 3 nodes\end{tabular} & \begin{tabular}[c]{@{}l@{}}$\chi^2_\mathrm{eff}$\\ $\ln B$\\ $\tau$\\ $\Delta^\mathrm{reion}_z$\end{tabular}
& \begin{tabular}[c]{@{}l@{}} -\\-\\-\\-\end{tabular}   
& \begin{tabular}[c]{@{}l@{}} 2781.8\\-6.6\\$0.05\pm0.005$\\$2.9^{+0.086+4.3}_{-0.82-2.1}$\end{tabular}          
& \begin{tabular}[c]{@{}l@{}} 2786.5\\-8.2\\$0.052^{+0.0006}_{-0.003}$\\$2.86^{+0.07+1.5}_{-0.6-0.86}$\end{tabular}               
\\ \hline
\end{tabular}
\caption{~\label{tab:bounds} Best fit $\chi^2_\mathrm{eff}=-2\ln {\cal L}$ from MCMC and the bounds on the optical depth $\tau$ (68.3\% C.L.) and duration of reionization $\Delta^\mathrm{reion}_z$ (both 68.3\% and 95\%C.L. for skewed posterior) obtained in the reconstructions for different data combinations. The Bayes factors ($\ln B$) {\it w.r.t.} the minimal model B0 are also provided.}  
\end{table}
\begin{figure*}
\resizebox{115pt}{88pt}{\includegraphics{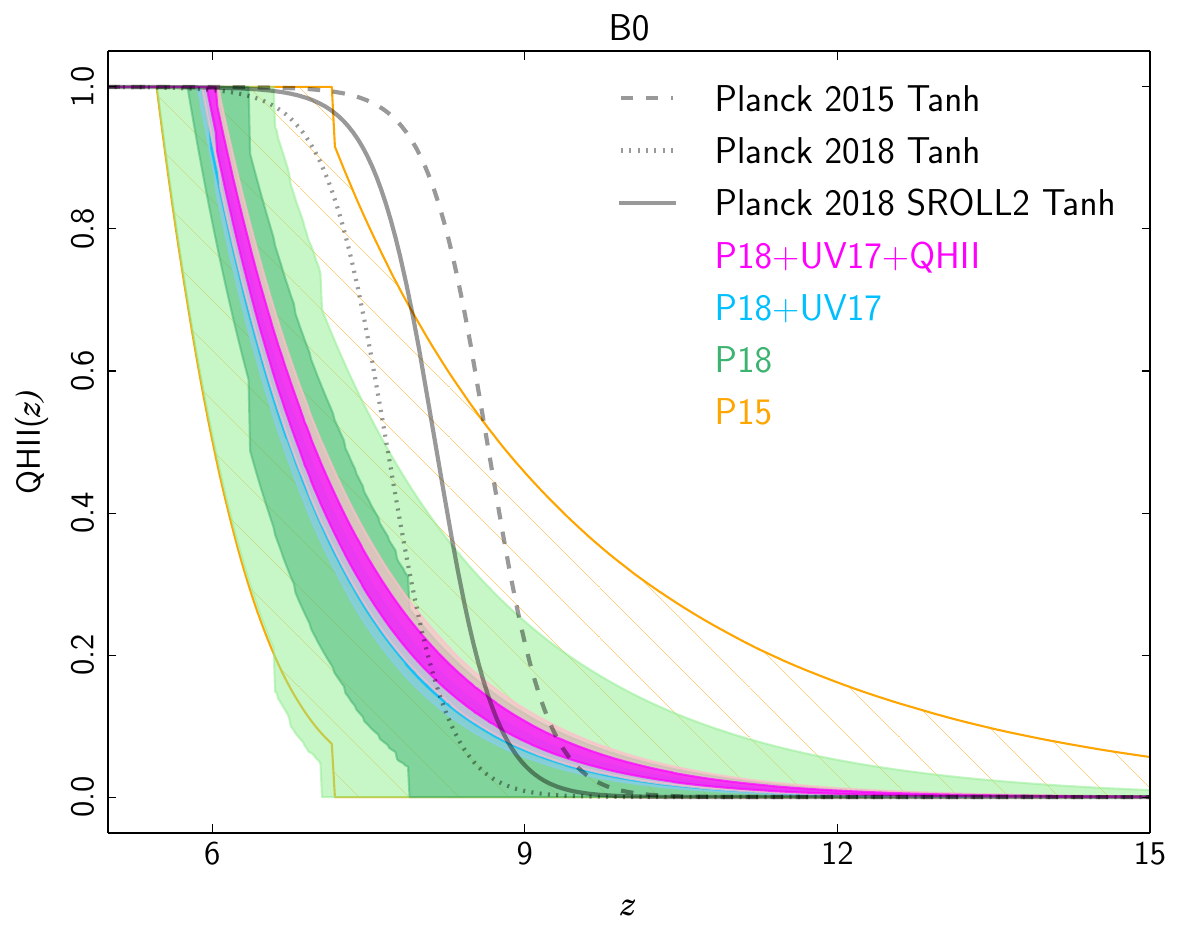}}
\resizebox{115pt}{88pt}{\includegraphics{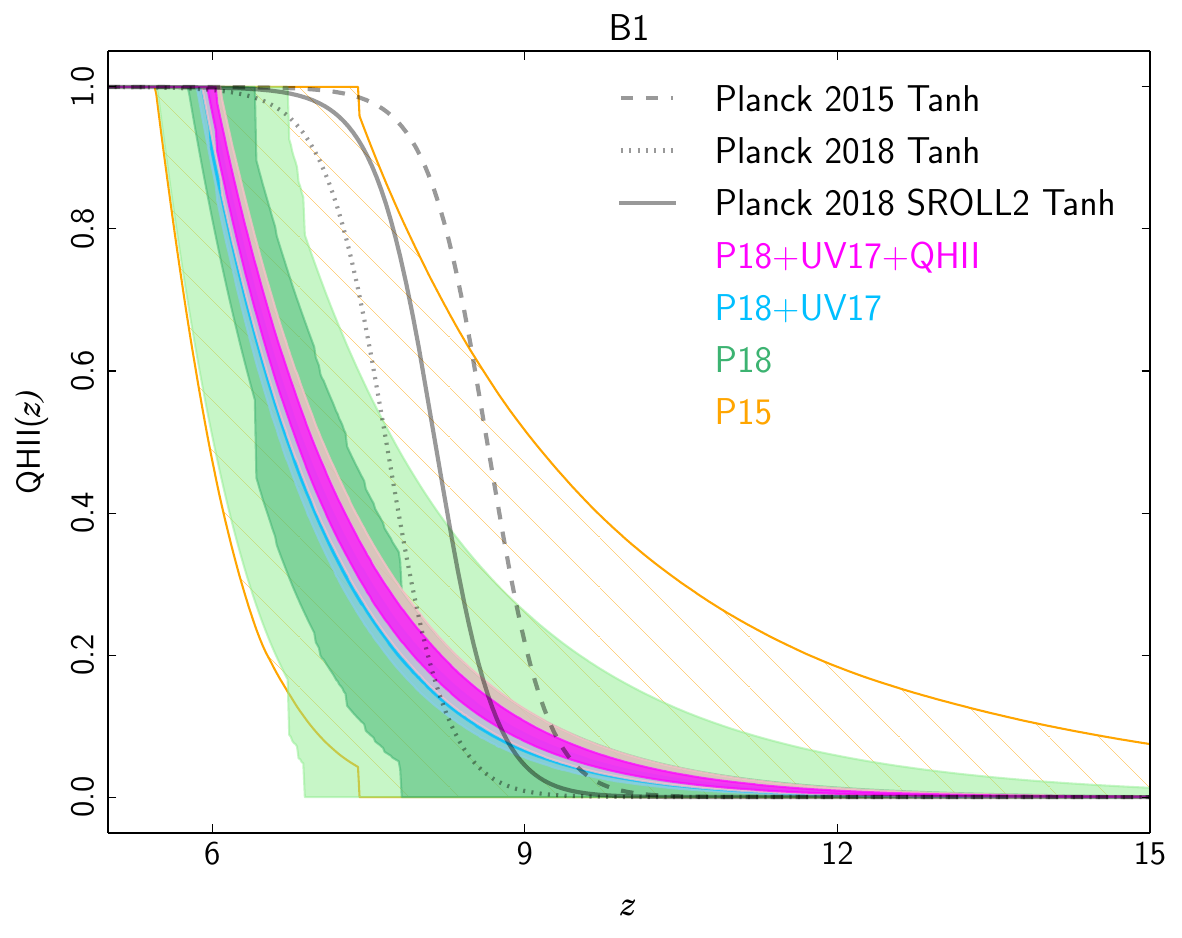}}
\resizebox{115pt}{88pt}{\includegraphics{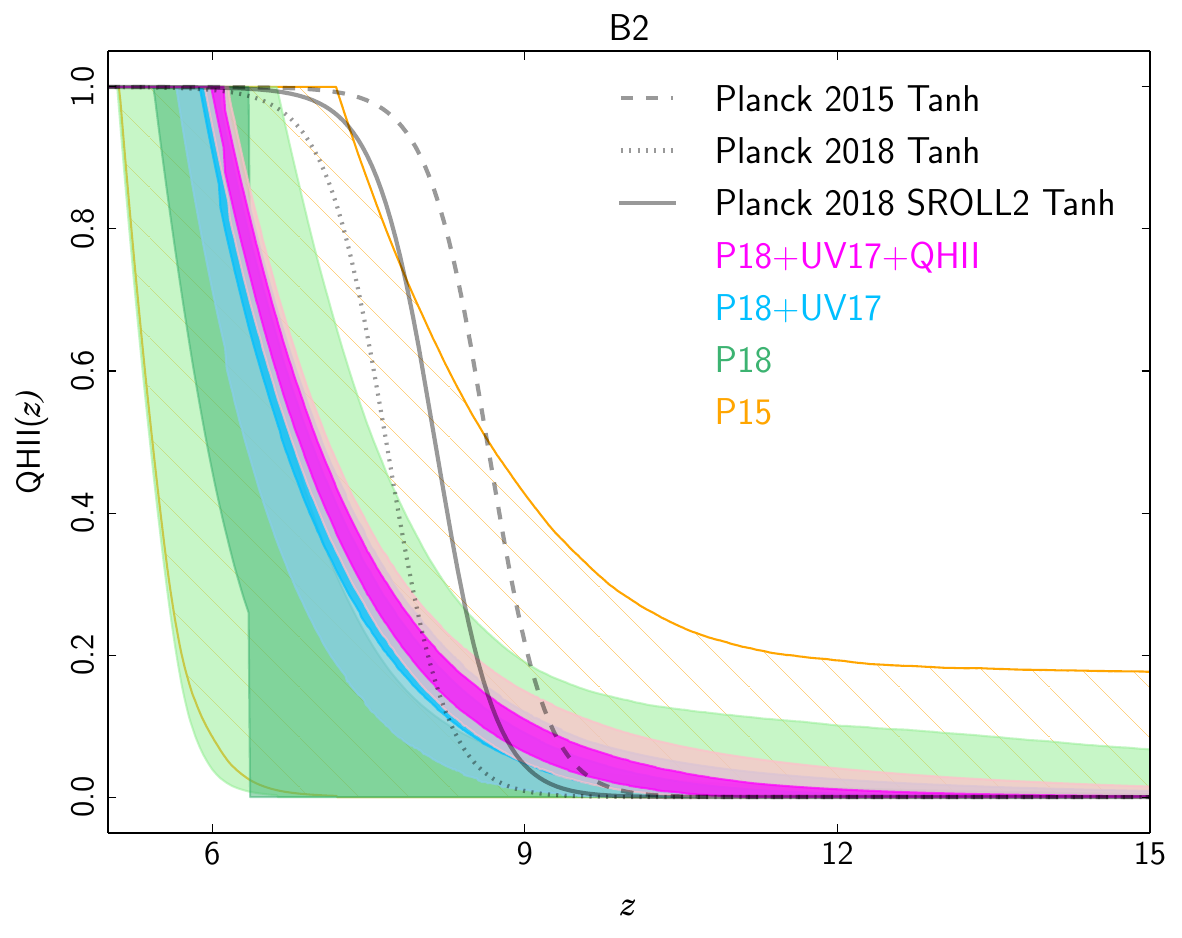}}
\resizebox{115pt}{88pt}{\includegraphics{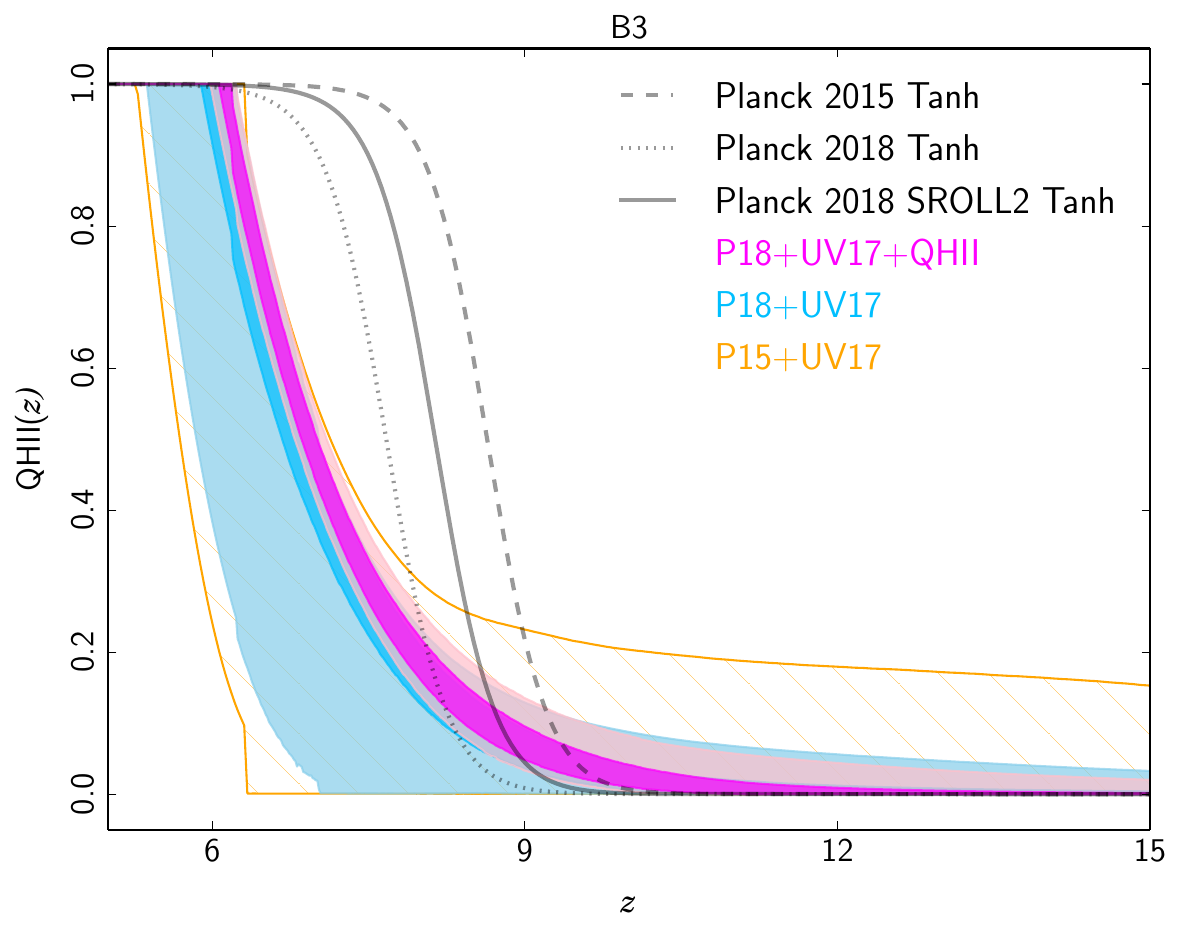}}

\resizebox{115pt}{88pt}{\includegraphics{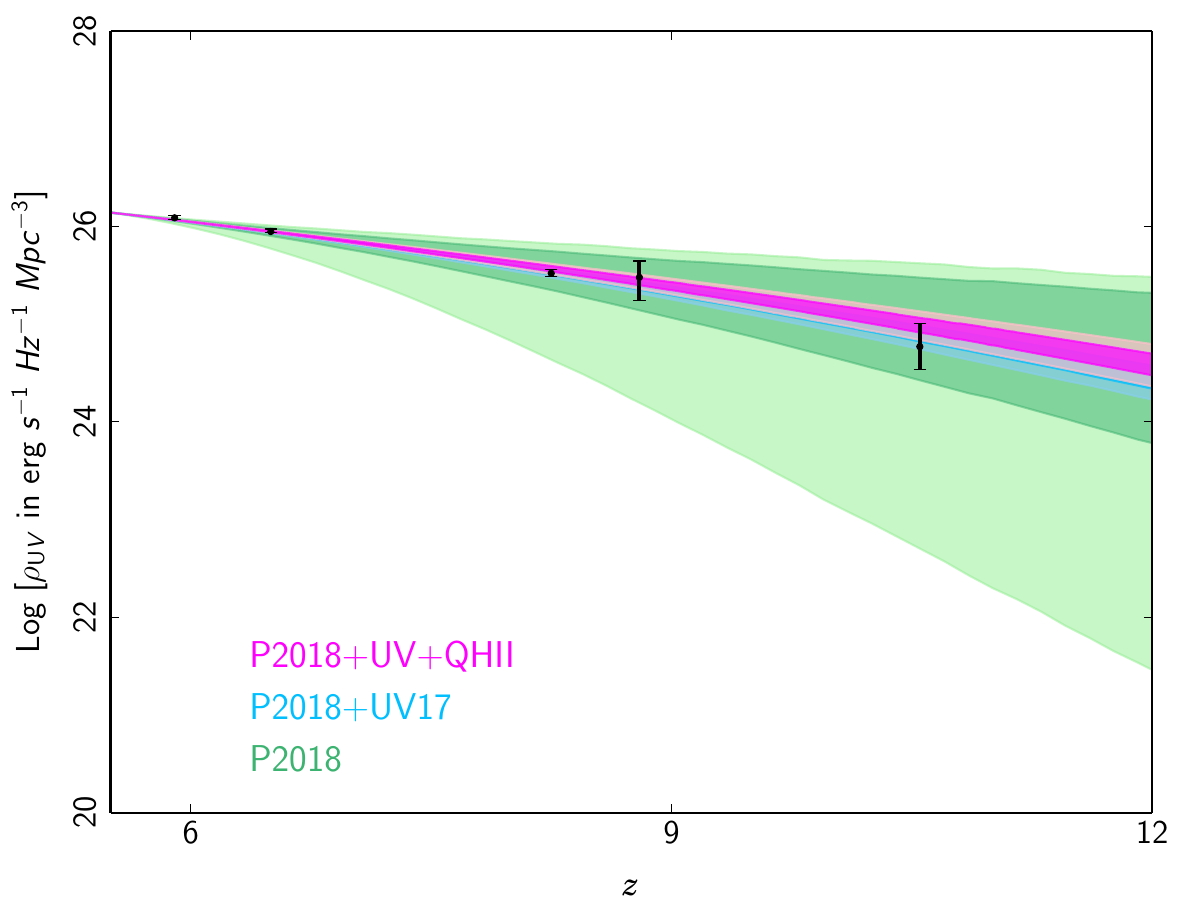}}
\resizebox{115pt}{88pt}{\includegraphics{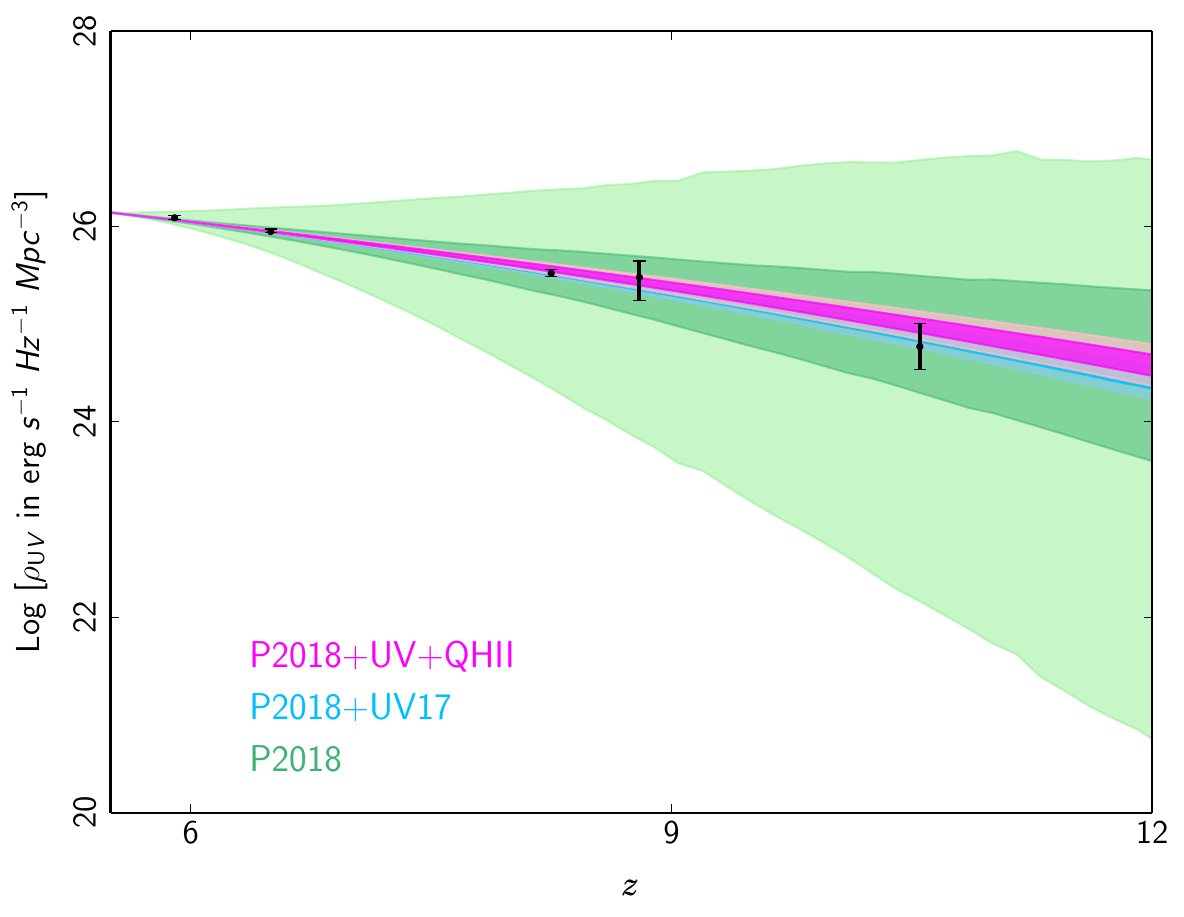}}
\resizebox{115pt}{88pt}{\includegraphics{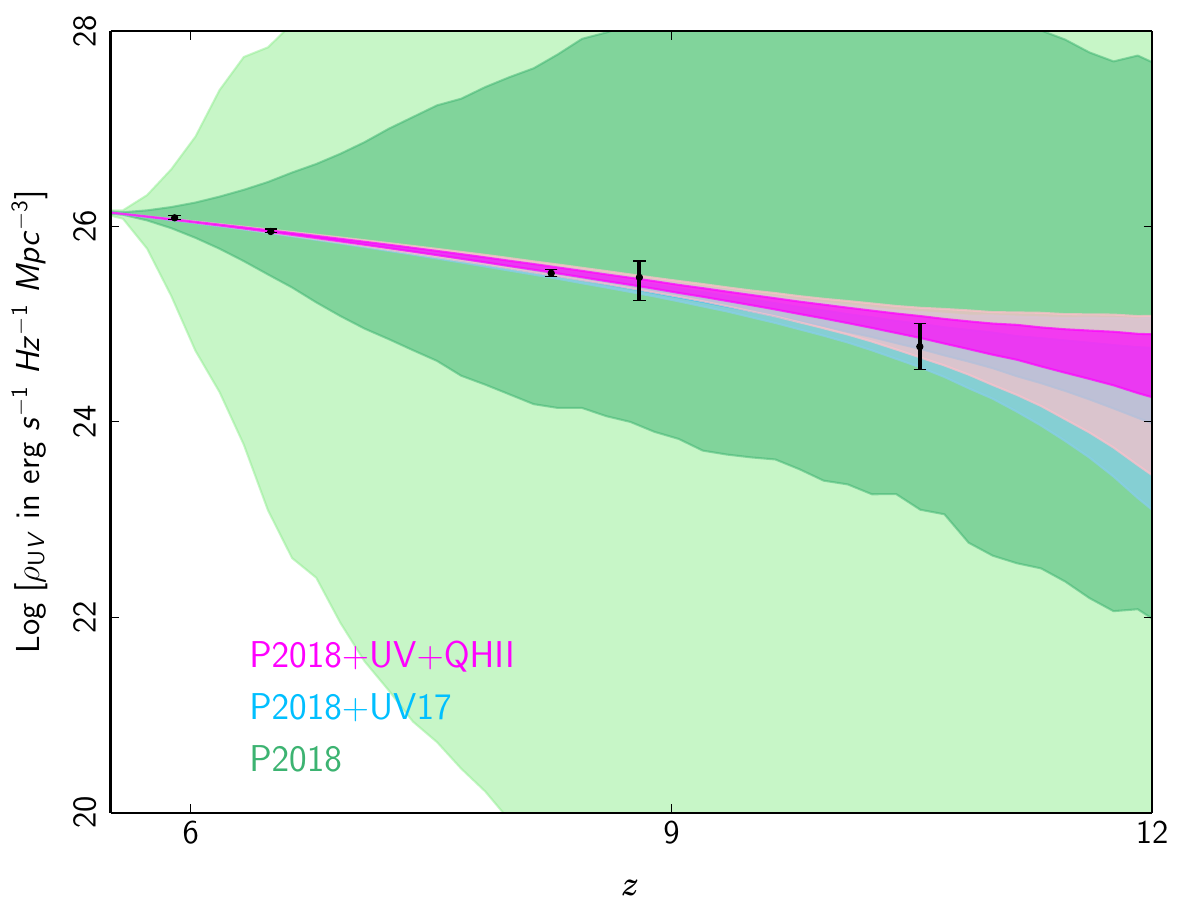}}
\resizebox{115pt}{88pt}{\includegraphics{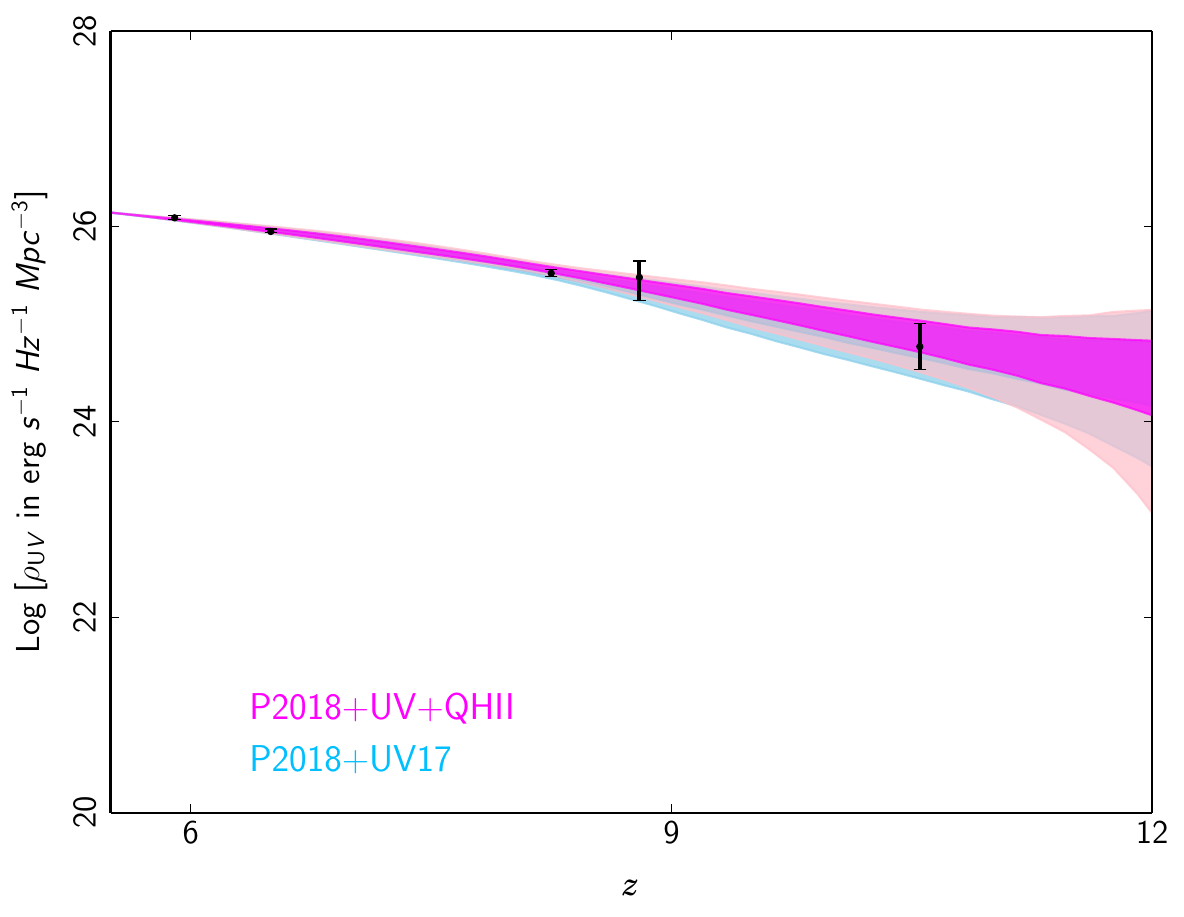}}

\resizebox{115pt}{88pt}{\includegraphics{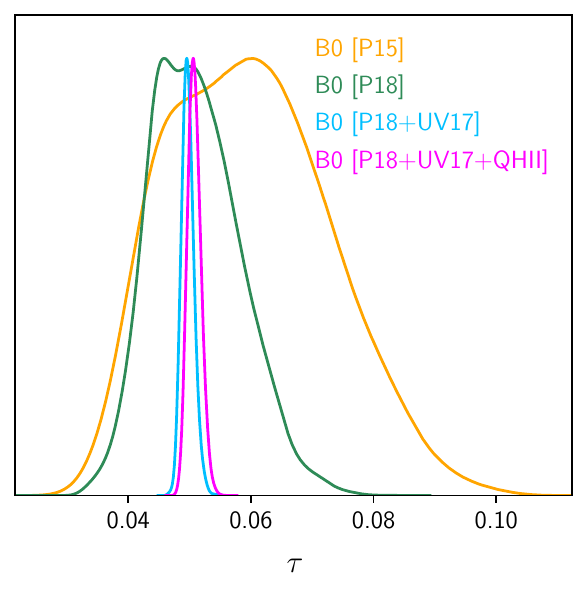}}
\resizebox{115pt}{88pt}{\includegraphics{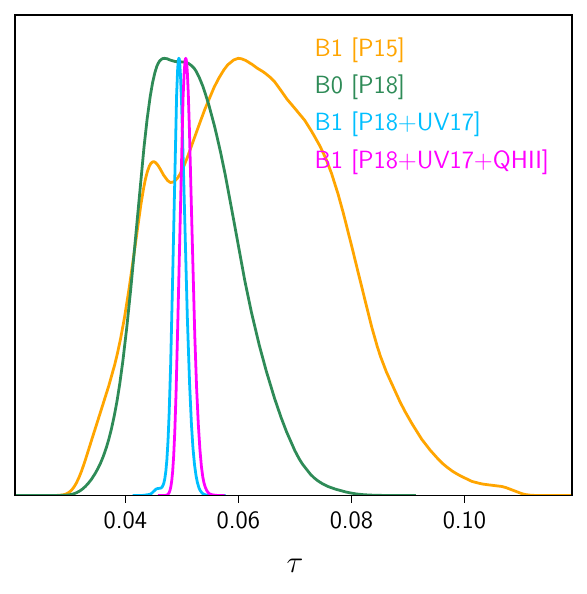}}
\resizebox{115pt}{88pt}{\includegraphics{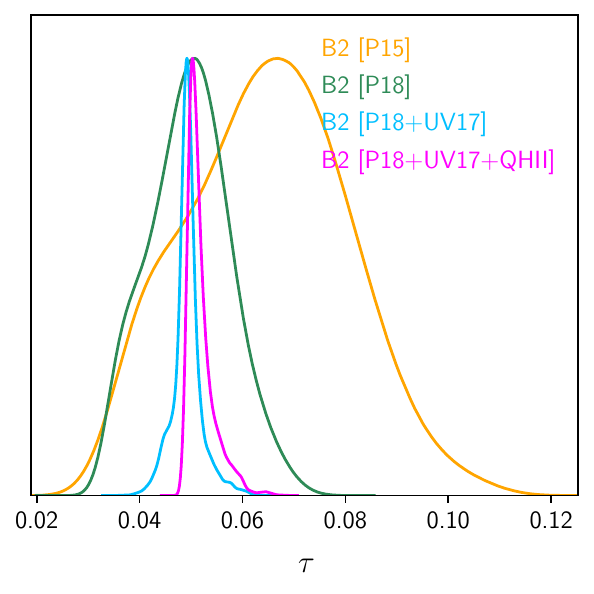}}
\resizebox{115pt}{88pt}{\includegraphics{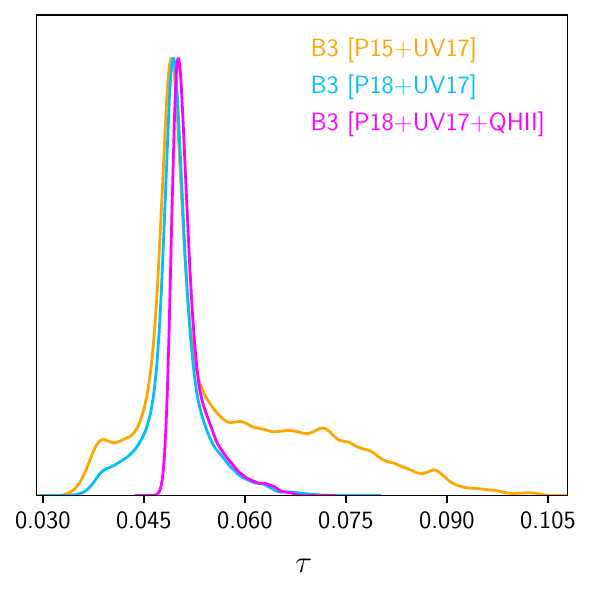}}
\caption{\label{fig:samples} [Left to right]: Results for minimal single node (B0), single node (B1), two nodes (B2) 
and three nodes (B3) reconstructions respectively. Planck best fits for Tanh reionization are plotted in grey. SROLL2 refers to the independent low-$\ell$ E-mode polarization likelihood based on Planck data~\cite{Delouis:2019bub}.
[Top]: The volume filling factor as a function of redshift. Constraints are computed from the entire MCMC samples. [Middle]: UV luminosity density with the Ishigaki {\it et. al.} (2018) compiled data (also containing Bouwens {\it et. al.} (2014)).  [Bottom]: Marginalized probability distribution function of $\tau$. It is evident from the plots that a sharp history of reionization can not make all three datasets agree.}  
\end{figure*}
\begin{figure}
\includegraphics[width=\columnwidth]{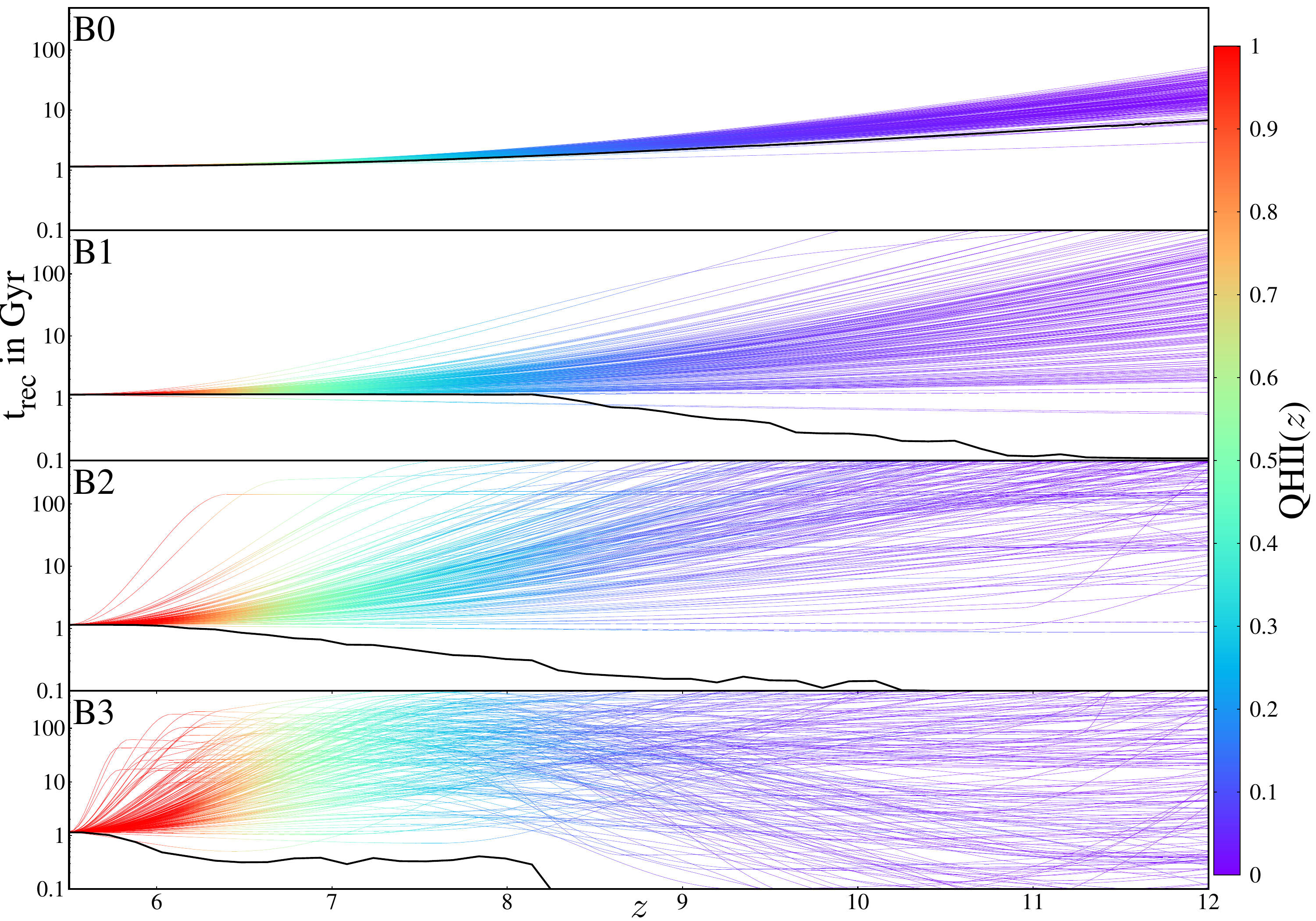}
\caption{\label{fig:trec-samples}Samples of recombination timescale in Gigayears as a function of redshift in minimal to three node cases (top to bottom) are plotted. The 2$\sigma$ lower bounds are also provided in thick black curves. The volume filling factors in all samples are colored to demonstrate the progress of reionization and its dependence on $t_\mathrm{rec}$.}  
\end{figure}
\begin{figure}
\flushleft\resizebox{240pt}{175pt}{\includegraphics{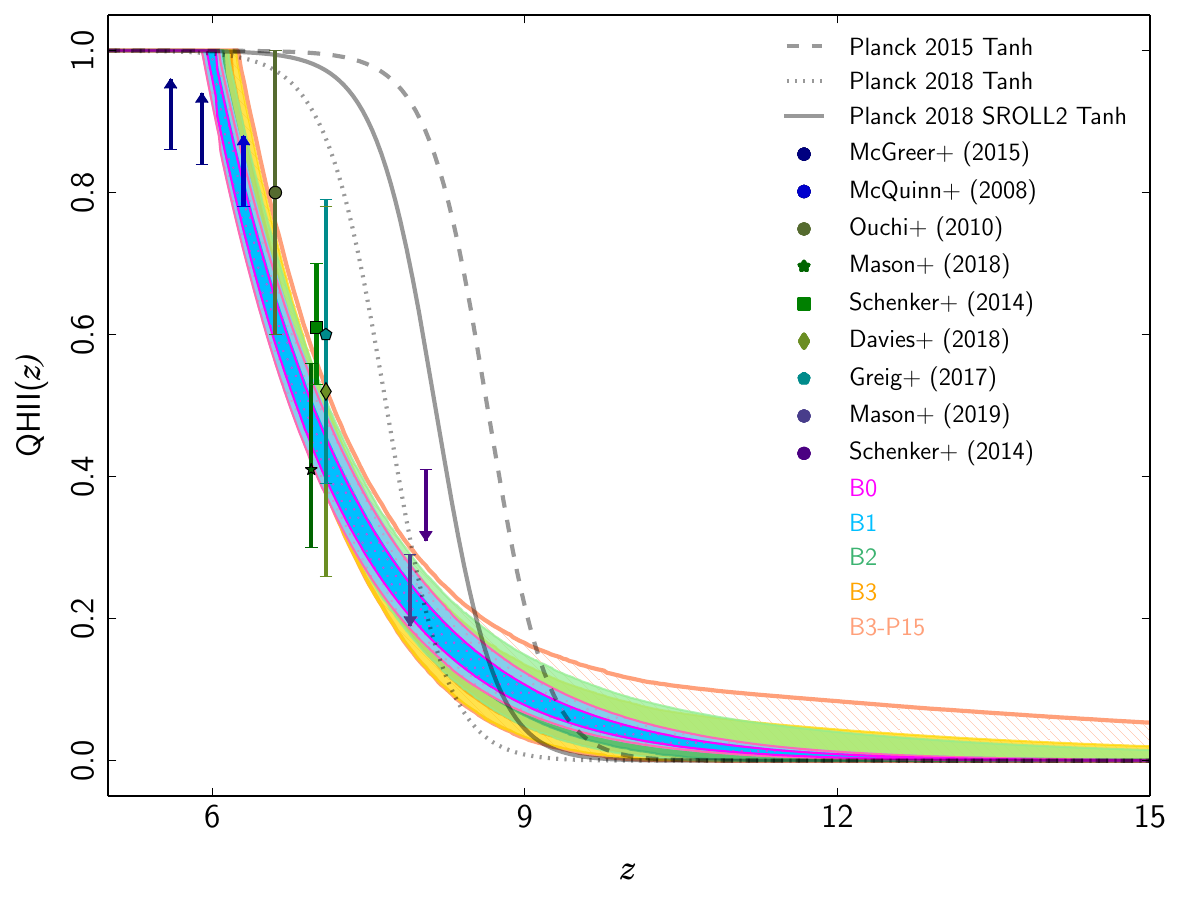}}
\caption{\label{fig:QHII} 68.3\% and 95\% C.L. on $Q_{\rm HII}$ as a function of redshift in four different cases considered. Data points and limits are also plotted. In grey dotted, solid and dashed we plot best fit Tanh model for Planck 2018, Planck 2018 with SROLL2 low-$\ell$ EE likelihood and Planck 2015 respectively. Family of reionization histories obtained in our reconstruction address the data more efficiently compared to Tanh model.}  
\end{figure}
\begin{figure}
\flushleft\resizebox{121pt}{121pt}{\includegraphics{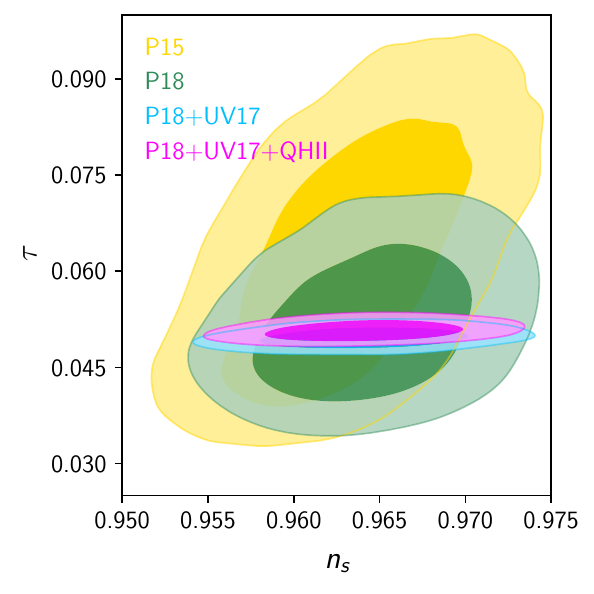}}
\resizebox{121pt}{121pt}{\includegraphics{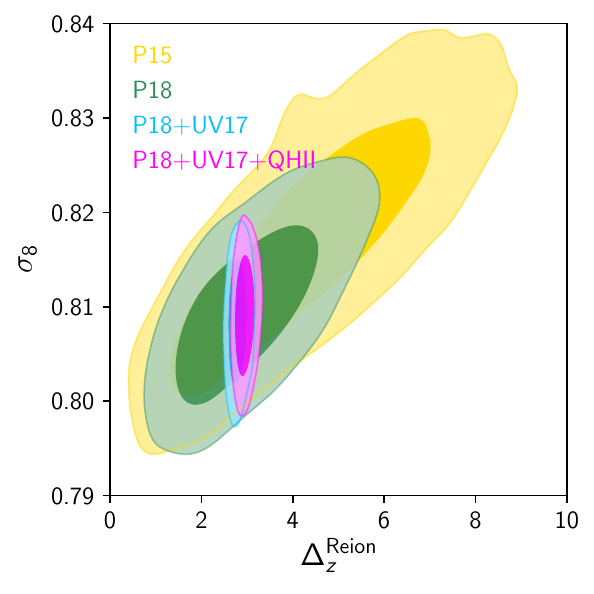}}
\caption{\label{fig:nstau} Correlations between the spectral index and the optical depth (left) and between the duration of reionization and $\sigma_8$ normalization (right) in single node (B1).}  
\end{figure}


In Table~\ref{tab:bounds} we provide the constraints on $\tau$, $\Delta^\mathrm{reion}_z$, 
i.e. the redshift interval between 10\% and complete ionization, the fit to the data ($\chi^2_\mathrm{eff}$) 
and the Bayesian evidences ($\ln B$) calculated from the chains using MCEvidence~\cite{Heavens:2017afc,Heavens:2017hkr}. We find that $\ln B$ for B1 is close to 0 {\it w.r.t} B0. While B2 and B3 improve the fit to all data combinations, the 
addition of extra parameters is penalized by the Bayes factor and become disfavored compared to B1.  It can be readily identified for B0 and B1, that allow monotonic histories, $\tau$ can be constrained with much better precision when our compilation of astrophysical data from UV17 and UV17+QHII are combined compared to P18 alone. In all the cases, mean values of $\tau$ remains similar and the low-$\ell$ polarization likelihood from P18 using HFI, plays an important role making the histories consistent with the astrophysical data. 
With more general histories allowed in B2, addition of UV17 and UV17+QHII only improves the constraint by 50\% compared to P18. Since in B2 and B3, the last node is only constrained by P18, upper bound on optical depth gets worse. In~\cite{Gorce2017}, keeping fixed the underlying cosmology and using five (four star formation history parameters and clumping factor) and six (allowing $f_{\rm esc}$ to vary alongside)
parameters, the authors report the standard deviations of $\tau$ to be $0.0019$ and $0.002$ respectively, when all datasets are used. 
In B2 and B3, that allows 6 and 9 parameters to describe the reionization, we obtain standard deviation$\sim0.003-0.0035$, 50\% wider compared to~\cite{Gorce2017}. The constraints become more conservative as our reconstructions allow more flexibilities compared to fixed form parametrization. While our framework allows a wide range of $\Delta^\mathrm{reion}_z$, the constraints demonstrate that in a P18+UV17+QHII data combination, even the most flexible model (B3) must have $\Delta^\mathrm{reion}_z>2$ at 95\% C.L. 

In Fig.~\ref{fig:samples} we plot the constraints from the MCMC analyses. We plot 95\% bounds on $Q_{\rm HII}$ (top row) for all the data combinations as a function of $z$. Constraints from P15 for B0-B2 and P15+UV17 for B3 are provided in the background stripes. The improvement in P18 constraints using HFI polarization compared to P15 data is significant, as anticipated in~\cite{Planck2016:HFIsys,Planck2016:reion}.  
While B0 produces monotonic power law reionization histories, B1 allows extended and step like histories. B2 and B3 with extra nodes provide the scopes for non-monotonic and complex histories. As we know, P18 mainly constrains the integrated optical
depth, therefore the ionization histories are not well constrained in all three cases (B0, B1 and B2). UV luminosity density data allow only a small subset of histories from P18 and the derived bounds on $\tau$ improve significantly in P18+UV17. In middle row, we plot 68\% and 95\% constraints on
corresponding source term, $\rho_{\rm UV}$ and on top we display the Bouwens {\it et. al.} (2014) and Ishigaki {\it et. al.} (2018) and data points used. Since luminosity densities at higher redshifts for B2 and B3 are not constrained well and therefore we plot samples 
only till $z=12$. 
In bottom row we plot the marginalized constraints on $\tau$: for all the cases the improvement due to our compilation of astrophysical data {\it w.r.t} CMB alone is evident. 
The optical depth from P18+UV17 and P18+UV17+QHII agree well in all cases. The agreement is better in P18 compared to P15 as the later inclines towards higher mean values of $\tau$, although with larger uncertainties.

In Fig.~\ref{fig:trec-samples} we plot reconstructed samples of recombination timescale and its lower limits obtained from P18+UV17+QHII.  Within $z=6-8$, the lower limit on the timescale is found to be about $1$ Gyr. The evolution of ionization is mapped with colored samples.   

In Fig.~\ref{fig:QHII}, we plot the 68.3\% and 95\% C.L. on the reionization histories for all the reconstructions using P18+UV17+QHII. On top of the bounds, we plot $Q_{\rm HII}$ data points from different observations that we have used. While B0 and B1 reconstruct similar histories, constraints on histories in B2 and B3 are wider at higher redshifts. Significant improvement compared to P15 is evident in B3.

Note that, in this framework we can also reconstruct the clumping factor at different redshifts. Since our sample provide free form reconstruction of $t_\mathrm{rec }$, we can obtain the clumping factor for any assumed value of IGM temperature.  For $T_\mathrm{IGM}=2\times10^4K$, we find $C_{\rm HII}\lesssim3$ within $6<z<8$ and monotonically increasing with decrease in redshift. This result is completely consistent with parametric $C_{\rm HII}=2.9\l[\frac{1+z}{6}\r]^{-1.1}$ fit to simulation~\cite{Shull2012}. This bound is expected to be degenerate with $\langle f_{\rm esc}\xi_{\rm ion}\rangle$ if allowed to be free (for more discussion on $f_{\rm esc}$ see,~\cite{Dayal2018}). 

Our analysis finds correlations between other background cosmological parameters with the reconstructed reionization histories and in Fig.~\ref{fig:nstau} we present the correlation between derived parameter $\tau$ and $n_{\rm s}$ and between $\sigma_8$ and $\Delta^\mathrm{reion}_z$ for B1. HFI polarization data on large angular scales and astrophysical data helps in breaking the degeneracies and provide tighter constraint on the reionization histories and therefore on $\tau$. Tighter constraints on $n_{\rm s}$ and $\sigma_8$ are also important to compare CMB with large scale structure data. 

\section{Conclusion}~\label{sec:conclusion} 
In this paper we have reconstructed the history of reionization using CMB and a compilation of astrophysical data. This free form reconstruction allows more conservative variation in ionizing UV flux (the source term $\dot{n}_{\rm ion}$) and recombination times/rates ($t_{\rm rec}$) at flexible redshift nodes. We are also able to combine data from UV luminosities in addition to those for CMB and neutral hydrogen. This framework allows sharp to highly extended reionization histories that also involves non-monotonic changes in the ionization fraction.

Below we summarize the main results of our analysis:

\begin{enumerate}[leftmargin=0.35cm]

\item  
We find an excellent consistency between the low-$\ell$ Planck 2018 HFI polarization likelihood and our compilation of astrophysical data in determining the integrated optical depth $\tau$. When considering jointly P18+UV17+QHII data, 
we obtain $\tau = 0.051^{+0.006+0.013}_{-0.009-0.014}$. We note that with the nominal errors 
of our compilation of astrophysical data, we obtain a joint constraint tighter by a nearly a factor of 2 {\it w.r.t.} the projected constraint from cosmic variance limited proposed CMB space missions~\cite{core,litebird,CMBBHARAT,pico}. 
 \item A joint analysis that includes Planck 2018 data, UV luminosity density integrated upto 
 -17 magnitude and Lyman-$\alpha$ observations, does not allow sharp reionization histories (with Tanh model defined as sharp, $\Delta^{\rm reion}_{z}\sim1.7$ between 10\% to 99\% ionization). We report at 95\% C.L. $2.6<\Delta^{\rm reion}_{z}<3.2$ (in the single node reconstruction), 
and $2<\Delta^{\rm reion}_{z}<4.4$ (three nodes reconstruction allowing conservative constraints).


 \item There are no evidences for non-monotonic or multi-step reionization histories. Bayesian evidence disfavors complex reionization histories with more than one intermediate node between the beginning and completion of reionization. Use of HFI large scale E-mode polarization in P18 results in substantially tighter constraints at the high redshift tail of ionization histories compared to P15.

 \item  Samples of recombination timescales from P18+UV17+QHII reveals that clumping factor $C_\mathrm{HII}\lesssim3$ within redshift $6-8$, assuming a IGM temperature of 20000K (correspondingly, we find 95\% lower bound of $t_\mathrm{rec}\sim1$ Gyr). 
 
\item When $\langle f_{\rm esc}\xi_{\rm ion}\rangle$ is allowed to vary, the combined datasets constrains $\tau=0.052\pm0.002$ at 95\% in the single node case. 

Allowing $\langle f_{\rm esc}\xi_{\rm ion}\rangle$ as free parameter use of different UV luminosity data with  truncation magnitude of -15 (UV15) provides $\tau=0.054\pm0.003$ at 95\% in single node case. Contribution towards ionization from dimmer sources is reflected in the higher value of optical depth. Higher uncertainty in the UV15 leads to relaxed bounds {\it w.r.t} UV17 case.  
 
Discussions on such extensions and constraints on  $f_{\rm esc}$, $\xi_{\rm ion}$  in different cases and data combinations will be provided in a detailed paper~\citep{HPFSinprep}. 
 
 \item We find that for simple monotonic models that can be described by a single intermediate node, degeneracies between reionization history and other cosmological parameters can be lifted completely with current astrophysical data. 
\end{enumerate}

Our analysis opens up to conservative constraints on reionization allowing the combination of astrophysical measurements with CMB in this newly introduced 
free-form reconstruction. It will be interesting to see the performance of such method for constraining physical models of reionization in the perspective of future cosmological measurements.

\begin{acknowledgments}
The authors would like to acknowledge the use of Nandadevi cluster in the Institute of Mathematical Science’s High Performance Computing (\href{hpc.imsc.res.in}{hpc.imsc.res.in}), APC cluster
(\href{https://www.apc.univ-paris7.fr/FACeWiki/pmwiki.php?n=Apc-cluster.Apc-cluster}{https://www.apc.univ-paris7.fr/FACeWiki/pmwiki.php?n=Apc-cluster.Apc-cluster}) 
and INAF OAS Bologna cluster.
The authors thank Masami Ouchi and Masafumi Ishigaki for providing their UV luminosity density data compilation.
DKH would like to thank Sourav Mitra for important discussions.
The authors would like to thank Andrea Ferrara and Tirthankar Roy Choudhury for their valuable comments on the manuscript.
DKH has received fundings from the European Union’s Horizon 2020 research and innovation programme under the Marie Sklodowska-Curie grant agreement  No. 664931.
DP and FF acknowledge financial support by ASI Grant 2016-24-H.0 and 
partial financial support by the ASI/INAF Agreement I/072/09/0 for the Planck LFI Activity of Phase E2.
GFS acknowledge Laboratoire APC-PCCP, Universit\'e Paris Diderot and Sorbonne Paris Cit\'e (DXCACHEXGS)
and also the financial support of the UnivEarthS Labex program at Sorbonne Paris Cit\'e (ANR-10-LABX-0023 and ANR-11-IDEX-0005-02). 
\end{acknowledgments}

\bibliography{reionref}

\end{document}